\documentclass[conference]{IEEEtran}
\IEEEoverridecommandlockouts
\usepackage{cite}
\usepackage{amsmath,amssymb,amsfonts}
\usepackage[linesnumbered,ruled]{algorithm2e}
\usepackage{graphicx}
\usepackage{float}
\usepackage{textcomp}
\usepackage{xcolor}
\usepackage{CJKutf8}
\usepackage{graphicx}
\usepackage{multirow}
\usepackage{color}
\usepackage{setspace}
\usepackage{makecell}
\usepackage{amsthm}
\usepackage{color}
\usepackage{indentfirst}
\usepackage{balance}
\usepackage{subcaption}
\usepackage{array}
\usepackage{flushend}
\usepackage{balance}
\usepackage{url}
\usepackage{booktabs}
\usepackage{ragged2e}
\usepackage{placeins}

\floatname{algorithm}{Algorithm} 
\newtheorem{myDef}{Definition}

\newtheorem{exmp}{Example}

\usepackage{comment}
\def\BibTeX{{\rm B\kern-.05em{\sc i\kern-.025em b}\kern-.08em
    T\kern-.1667em\lower.7ex\hbox{E}\kern-.125emX}}

\begin{document}
\title{Time Sensitive Multiple POIs Route Planning on Bus Networks}
\author{  
\IEEEauthorblockN{  
Simu Liu\IEEEauthorrefmark{1},  
\textcolor{black}{Kailin Jiao}\IEEEauthorrefmark{2},  
\textcolor{black}{Junping Du}\IEEEauthorrefmark{3},  
\textcolor{black}{Yawen Li}\IEEEauthorrefmark{3},  
\textcolor{black}{Zhe Xue}\IEEEauthorrefmark{3},  
\textcolor{black}{Xiaoyang Sean Wang}\IEEEauthorrefmark{4},  
\textcolor{black}{Ziqiang Yu}\IEEEauthorrefmark{1*}\thanks{*Corresponding author},
Yunchuan Shi\IEEEauthorrefmark{1}\\}  

\IEEEauthorblockA{  
\IEEEauthorrefmark{1}Yantai University \IEEEauthorrefmark{2}Shandong University \IEEEauthorrefmark{3}Beijing University of Posts and Telecommunications \IEEEauthorrefmark{4}Fudan University}  
}
\DeclareRobustCommand*{\IEEEauthorrefmark}[1]{%
    \raisebox{0pt}[0pt][0pt]{\textsuperscript{\footnotesize\ensuremath{#1}}}}

\maketitle

\begin{abstract}
\begin{CJK*}{UTF8}{gbsn}
This work addresses a route planning problem constrained by a bus road network that includes the schedules of all buses. Given a query with a starting bus stop and a set of Points of Interest (POIs) to visit, our goal is to find an optimal route on the bus network that allows the user to visit all specified POIs from the starting stop with minimal travel time, which includes both bus travel time and waiting time at bus stops. Although this problem resembles a variant of the Traveling Salesman Problem, it cannot be effectively solved using existing solutions due to the complex nature of bus networks, particularly the constantly changing bus travel times and user waiting times. In this paper, we first propose a modified graph structure to represent the bus network, accommodating the varying bus travel times and their arrival schedules at each stop. Initially, we suggest a brute-force exploration algorithm based on the Dijkstra principle to evaluate all potential routes and determine the best one; however, this approach is too costly for large bus networks. To address this, we introduce the EA-Star algorithm, which focuses on computing the shortest route for promising POI visit sequences. The algorithm includes a terminal condition that halts evaluation once the optimal route is identified, avoiding the need to evaluate all possible POI sequences. During the computation of the shortest route for each POI visiting sequence, it employs the A* algorithm on the modified graph structure, narrowing the search space toward the destination and improving search efficiency. Experiments using New York bus network datasets demonstrate the effectiveness of our approach.
 \end{CJK*}
\end{abstract}


\section{Introduction}\label{sec:intro}
\begin{CJK*}{UTF8}{gbsn}

In modern urban areas, public transportation offers an efficient and cost-effective alternative to taxis. When a tourist visits an unfamiliar city and wishes to explore multiple Points of Interest (POIs) using buses, they often require a route planning service. An ideal route planning service allows them to specify a starting bus station and the POIs they wish to visit, after which an optimal bus route can be provided that minimizes travel time. This route acts as a bus schedule, detailing which buses to board at each stop to reach the bus stations closest to the desired POIs starting from the original station. The travel time for this route includes both the riding and waiting times, which usually varies with the changing traffic condition. We refer to this ideal route planning service as the Time Sensitive Multi-POIs (TSMP) route planning problem on a bus network.

At first glance, the bus network could be represented as a graph, making this problem a variant of the Traveling Salesman Problem (TSP). However, this query is much more complicated than TSP due to the varying bus travel time and the time of waiting buses. More worse, the bus travel time and the waiting time differ for each bus and each station. Consequently, existing route planning methods~\cite{r18,r19,r20,r24,r25,r26,chen2018price,ding2018ultraman} based on general graphs cannot directly address ISMP queries within the bus network context.


To tackle this issue, we employ a Directed Multiple-Edges Graph, referred to as a DME-Graph, to model the bus network. In this representation, each bus station is depicted as a vertex, while the connections between adjacent bus stations are represented as edges. Given that there are various bus routes serving the same pair of stops, multiple edges exist between two adjacent vertices. In this graph, both vertices and edges are assigned weights. The weight of a vertex consists of a list of $\lbrace key, value\rbrace$ pairs, with each pair reflecting the arrival time of a bus at that station; the key corresponds to the bus identifier, and the value represents the arrival time. It’s important to note that these arrival times are continuously updated as buses operate. The weight of each edge indicates the average travel time between the two nearby bus stations, which can be derived from historical data.

Using the DME-Graph as a foundation, we introduce a novel method for identifying the shortest path. This involves enumerating all sequences made up of unordered points of interest (POIs), breaking each sequence down into multiple Origin-Destination (OD) pairs, and applying Dijkstra's algorithm to find the shortest path for each OD pair.  The final step involves comparing these results to identify the optimal overall route. However, Dijkstra's algorithm explores the DM-Graph in all directions from the starting point, which can lead to excessive and unnecessary searching when the destination is constrained to one direction. To address this issue, heuristic algorithms such as A*~\cite{r16,r17,r21,r22,r23,r30} and its various adaptations~\cite{r6,r8,r9,r10,r12,r13,r31,r32,r33,r38, r36,r15,r34,r14, r35} are proposed to enhance the efficiency of shortest path identification. Unfortunately, many of these algorithms struggle with the unique characteristics of the DM-Graph, which differs significantly from standard graphs. Additionally, while some methods achieve quicker point-to-point route identification by creating an index on the road network \cite{r39,r40,r33}, they are typically not suitable for the DM-Graph due to the dynamic weights, which can quickly render the path index ineffective.
\\
\indent To tackle the ISMP route planning challenges on the DM-Graph, we introduce an A* algorithm based on Euclidean distance, referred to as EA-Star. When expanding at each station, EA-Star predicts the minimal arrival time by computing the Euclidean distance from the station to the next POI and incorporates this with the weights of the DM-Graph to establish an estimated cost. This estimated cost helps to identify the most promising direction toward the next POI during the exploration process, and narrow the search area.

For a given TSMP query, where multiple POIs can be visited in any order, there are numerous potential visitation sequences, each corresponding to a possible route. Therefore, it is crucial to effectively prune unpromising POI visitation sequences to expedite the identification of the optimal route among the exponentially large number of potential options. To achieve this, we propose a pruning strategy based on Euclidean distance. This involves calculating the Euclidean distance for each visitation sequence and ranking these sequences accordingly. We then prioritize the processing of sequences with smaller Euclidean distances and establish a score threshold based on the shortest time identified thus far. For each sequence under consideration, we estimate its maximum score, factoring in the Euclidean distance and the maximum bus driving speed. If the estimated maximum score falls below the current score threshold, that sequence—along with those ranked behind it—can be safely disregarded.
\\
\indent In summary, the contributions of this work are as follows:
\\ \indent (1) We present the Time Sensitive Multiple POIs (TSMP) route planning query specifically for bus  networks. This query differs significantly from typical route planning in standard road networks due to the distinct features of bus systems. It has the potential to significantly improve the user experience of bus services.
\\ \indent (2) We introduce a Directed Multiple-Edges Graph (DME-Graph) to effectively represent the bus network. This model accurately captures the various bus routes linking to bus stations and their respective schedules, creating a solid foundation for addressing TSMP queries.
\\ \indent (3) We present EA-Star to address TSMP queries. It leverages the Euclidean distances of potential routes to preemptively filter out less promising POI visiting sequences, avoiding the exhaustive computation of exact travel times for every possible sequence of POI visits.
\\ \indent (4) Extensive experiments are conducted using datasets derived from the New York bus network to verify the performance of our proposed solution.
\\
\indent The paper is organized as follows: Section~\ref{Sec:related work} discusses the related work, while Section~\ref{Sec:problem definition} provides essential definitions. Section~\ref{naive approach} details the first attempt using the DSRS algorithm. and Section~\ref{Sec:eastar} introduces the EA-Star algorithm. Section~\ref{Sec:expertiments} outlines the results of our experimental evaluations, and Section~\ref{Sec:conclusion} concludes the paper.
\end{CJK*}

\section{Related Work}\label{Sec:related work}
\begin{CJK*}{UTF8}{gbsn}
Route planning problems on road networks have been extensively researched. Based on route search strategies, these problems can be broadly classified into traditional search algorithms and heuristic search algorithms. We will investigate the specific solutions currently available for these aforementioned two categories.

\indent\textbf{Incremental Expansion Algorithm}.
Incremental Expansion Algorithm\cite{r18}, particularly Dijkstra, play a pivotal role in determining the shortest path\cite{r5}. Additionally, a proposed MOD algorithm designed for multiple source points and a single destination is also utilized in public transportation network planning\cite{r7}. In contrast to the focus of this paper, we address the challenge of multiple POIs with single source points. Moreover, Dijkstra can be integrated with other algorithms. A combination of Dijkstra and Floyd-Warshall algorithms\cite{r20} , known for their best path selection, has been suggested for small-scale planning\cite{r24,r25,r26}. However, these algorithms are insufficient to address challenges in road networks, including destination disorder and dynamic time-based edge weights related to bus transfers and travel.The drawbacks of the shortest path algorithm were analyzed, and an improved version of Dijkstra's algorithm was proposed \cite{r6}. Since Dijkstra's algorithm is not suitable, a new algorithm based on the minimum number of transmissions was proposed\cite{r8}. Additionally, an optimal path in k-shortest paths is proposed considering user preferences\cite{r9}. A modified version of Dijkstra's algorithm\cite{r10}, focusing on transfer and walking distance, offers a more rational approach to route planning. In our work, we remain centered on finding the shortest path based on Dijkstra's algorithm, but emphasizes factors such as transmission, transfers, and user preferences\cite{r8,r9,r10}, rather than solely minimizing passengers' travel time.
\\
\indent \textbf{Heuristic algorithms}. Heuristics play a crucial role in route planning, the A-Star algorithm can expedite the search significantly\cite{r28,r29}.  These algorithms are commonly utilized in intricate scenarios like robot path planning, UAV path planning, and vessel vehicle path planning, among others\cite{r12,r13,r31,r32,r33,r38}. However, Our objective is to address the challenge of conducting travel time-based shortest path searches using the A-Star algorithm on the public transportation network.
\\
\indent Additionally, indexing methods such as IBAS and g-tree have been introduced to accelerate path recognition.\cite{r33,r39,r40}, but frequent updating of road network edge rights will make the index information inaccurate, increase the difficulty of index maintenance, and is not conducive to real-time path planning. Tang et al. introduced a data-driven multi-objective programming model that optimizes bus schedules by considering travel and waiting times, contrasting with our focus on minimizing passengers' total time spent.
A learning-based route selection method is proposed to address crime risks and reduce distances\cite{r36}. Researchers suggest extending the Ant Colony Optimization algorithm for finding the shortest route in mobile crowd sensing environments\cite{r15}. Night bus route planning focuses on identifying optimal unidirectional routes for specific origin-destination pairs, considering asymmetric passenger flows and enhancing stop selection algorithms using a two-way probabilistic propagation approach \cite{r34}. Our study shares similarities with works on shortest path finding \cite{r35}, but does not involve bidirectional route selection.
\\
\indent In summary, although path planning for road networks has been extensively studied, several problems have arisen from previous work in solving this problem: \textit{1)} some algorithms do not take into account the limitations of localized search during path search, leading to poor search efficiency on large road networks; \textit{2)} some road networks are not suitable for calculating the dynamic time cost, and do not consider the real-time waiting time well, which in turn affects the accuracy of the shortest paths; \textit{3)} The lack of sufficient consideration of POIs disorder leads to exponential growth of candidate sequences.



\end{CJK*}

\section{Problem Definition}\label{Sec:problem definition}
\begin{myDef}[DME-Graph] 
The Directed Multiple-Edges Graph $DG = (V,E,W)$ satisfies \textit{1)} a vertex $v_i \in V$  represents a bus stop with a
geographical coordinate (longitude, latitude); 
\textit{2)} an edge $e_{i,j,k}\in E$ represents a bus route between stops (i.e., vertices) $v_i$ and $v_j$, where $k$ denotes the identifier of the specific bus corresponding to this route; \textit{3)} both any edge $e_{i,j,k}$ and any vertex $v_i$ have an associated weight. The weight $w_{i,j,k}$ of the edge $e_{i,j,k}$ denotes the average riding time of the bus route between vertices $v_i$ and $v_j$, while $w_i$ of the vertex $v_i$ is a list of $\langle key, value\rangle$ pairs. Each $\langle key, value\rangle$ pair records the time schedule of a bus arriving at the vertex $v_i$, where ``key'' is the identifier of the bus and ``value'' is a sequence of timestamps indicating the bus's arrival times at this vertex.
\end{myDef}

In reality, each bus follows a bus line that comprises multiple stops. The collection of all stops across all bus lines makes up the complete set of vertices in the DME-Graph. If two bus stops are next to each other on at least one bus line, they are referred to as adjacent bus stops. There can be multiple bus routes connecting two adjacent bus stops, which results in multiple edges between stops in the DME-Graph.

\begin{CJK*}{UTF8}{gbsn}

\begin{figure}[!htbp]
\hspace{1.0cm}
    \includegraphics[width = 0.4\textwidth]{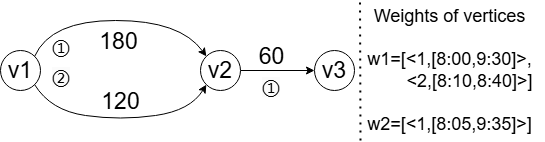}
    \caption{DME-Graph Instance}
    \label{net}
\end{figure}

\begin{exmp}
\label{dme}
 On the New York bus network, the bus ``Bx10'' passes through the stops ``PAUL AV/W MOSHOLU PY S'' and ``PAUL AV/W 205 ST'', while the bus ``Bx28'' passes through ``PAUL AV/W MOSHOLU PY S'', ``PAUL AV/W 205 ST'', and ``SEDGWICK AV/DICKINSON AV''. We use  $v_1$, $v_2$, and $v_3$ to denote ``PAUL AV/W MOSHOLU PY S'', ``PAUL AV/W 205 ST'', and ``SEDGWICK AV/DICKINSON AV'' respectively. The buse lines "Bx10" and "Bx28" are identified by the IDs ``1'' and ``2''. This segment of the bus network is represented as a graph shown in Fig.~\ref{net}. The weight of $v_1$, noted as $w_1$ = $[\langle 1, [8:00, 8:30] \rangle, \langle 2, [8:10, 8:40] \rangle]$ and $w_1$ = $[\langle 1, [8:05, 9:35] \rangle]$ means that the bus 1 will arrive at stop $v_1$ at 8:10 and 8:30, while the bus 2 will arrive at 8:00 and 8:30. The weight of edge $e_{1,2,1}$, given as $w_{1,2,1}=180$, indicates that the bus 1 will take 180 seconds driving from $v_1$ to $v_2$. 
\end{exmp}
\end{CJK*}

Due to multiple routes between two adjacent stops, the specific route between any two vertices must be defined not only by the sequence of vertices but also by the edges traversed. Therefore, we define the route as follows.

\begin{myDef}[Route] 
In the DME-Graph $DG$, for any two vertices the origin $v_s$ and the destination $v_t$, we choose any route denoted as $r(s,t)$ from $v_s$ to $v_t$ that consists of a set of $\lbrace \langle v_s,e_{s,a,k_x}\rangle, \cdots, \langle v_i,e_{i,j,k_y}\rangle, \cdots, \langle v_{l}, e_{l,t,k_z}\rangle\}$, where the bus lines ($k_x$, $\cdots$, $k_y$, $\cdots$, $k_z$) are permitted to be the same.

The route time cost, denoted by $C(r(s,t))$ consists of two components: \textit{1)} the sum of the edge weights $w_{i,j,k}$ along the route $r(s,t)$, which represents total riding time. \textit{2)}  The sum of the waiting time for each vertex $v_i$, derived from the current time and the weight of each vertex along  $r(s,t)$, represents the total waiting time.
\end{myDef}

\begin{myDef}[Minimum Cost]  
For any two vertices \( v_s \) and \( v_t \) in DME-Graph \( DG \), there are multiple routes \( r_i(s,t) \). The minimum cost of these routes, denoted by \( MC(s,t) \), is the smallest value among all route time costs between $v_s$ and $v_t$. The specific route between $v_s$ and $v_j$ with the minimum cost is denoted as $r_{min}(s,t)$.

\end{myDef}

\begin{myDef}[Candidate Route]
Given an origin vertex $v_s$ and a set of 
POIs $P=\{v_1, v_2,\ldots, v_n\}$, there exist multiple POI visiting sequences. Each sequence $SP_i = \langle v_0 = v_s, ..., v_n = v_l\rangle$ indicates an order of visiting POIs in $P$ and corresponds to a candidate route $CP_i$ passing through $v_s$ and all POIs along $CP_i$. 
\end{myDef}

A POI visiting sequence can correspond to multiple routes connecting $v_s$ and consecutive POIs by choosing different edges between adjacent vertices, where the route with the minimum time cost is referred to as {\em candidate route} for this sequence. For a candidate route $CP_i$,  its time cost is represented as $MC(CP_i) = \sum_{l=0}^{l\textless n} MC(l, l + 1)$.

\begin{myDef}[TSMP Route Query] 
In the DME-Graph $ DG(V,E,W) $, the TSMP route query $ q(v_s,t_{cur},P,DT) $ refers to the route with the minimum time cost among all candidate routes, denoted by $CP_{min}$. Here, $t_{cur}$ represents the current time, $ P $ denotes a set of POIs, and $ DT $ represents the set of dwell times $ dt $ for each $ v_i $ in $ P $. The route $CP_{min}$ adheres to the condition that for any candidate route $ CP_i $, $ MC(CP_{min}) \leq MC(CP_i) $ for $ m \neq i$.

\end{myDef}


\section{A First Attempt}
\label{naive approach}
The main challenge of the TSMP route planning query is  the potentially large number of candidate routes that arise from the set of Points of Interest (POIs) and the dynamic nature of bus arrival times. This variability allows for multiple order permutations when planning the route. Our initial algorithm involves enumerating all possible candidate routes and calculating the time cost for each one to ultimately identify the shortest route.

For a query specifying $n$ POIs, there are $n!$ POI visiting squences, each corresponding to a candidate route. To identify the candidate route, we decompose the sequence into multiple Origin-Destination (OD) pairs and apply Dijkstra's algorithm to determine the route with the minimum cost for each OD pair. These computed routes collectively form the candidate route as per Definition 4. Finally, we evaluate all $n!$ candidate routes to identify the route with the minimum time cost. Next, we will concentrate on the identification of the minimum time cost for an OD pair and the corresponding  route.

\subsection{Shortest route search for OD pairs}
\label{Sec:Shortest route search for OD pairs}
Given a candidate sequence $SP_i=\langle v_0, ..., v_l, ...,v_n\rangle$. We divide it into OD pairs $(v_0,v_1), ..., (v_l,v_{l+1}), ..., (v_{n-1},v_n)$.
Next, we propose a Dijkstra-based Shortest Route Search (DSRS) algorithm to identify the shortest route for each OD pair $(v_l,v_{l+1})$, and the pseudo-code is shown in Algorithm~\ref{alg:DSRS}. First, we initialize the time cost \( c \) of all vertices in the DME-Graph $DG$ to infinity, set the time cost \( c_l \) of the starting vertex \( v_l \) to zero, and mark \( v_l \) as visited (Lines 1-6).  

Next, we determine the waiting time for different bus routes starting from \( v_l \). For each subsequent bus stop $v_x$, it records a pair $\langle stop,route \rangle$ indicating that an arrival bus route from a stop to $v_x$, denoted by $pre_x$. For each adjacent vertex $v_j$ of $v_l$ ($v_j\in N(v_l)$), we first compare whether {$pre_i$} is the same as the bus route of the edge $e_{l,j,k}$. If it is the same bus route, we do not need to compute the waiting time at $v_j$. Otherwise, for each edge from \( v_l \) to the vertices in \( N(v_l) \), we match the specific bus line key \( k \) with \( w_l \) to obtain a sequence of timestamps indicating the arrival times of the bus line $k$. We can identify the nearest bus arrival time to the current time $t_{cur}$ based on these timestamps of $w_i$, denoted by \( wt(t_{cur},w_i) \) allowing us to compute the waiting time as the difference between $t_{cur}$ and the arrival times of the vertex $v_i$. Therefore, the time cost is the sum of the waiting time and the edge weight, expressed as \( c_j = wt(t,w_i) + w_{l,j,k} + c_l \). The current time $t_{cur}$ is updated upon arriving at the next vertex. Next, we select the vertex with the minimum time cost from the unvisited vertices as the vertex \( v_i \) to be visited, and repeat the process of updating the time cost for \( N(v_i) \) until all vertices have been visited or the minimum cost from $v_l$ to the target vertex $v_{l+1}$ has been determined (Lines 7-17). 

Ultimately, we construct the shortest route by backtracking the arrival bus route from $v_{l+1}$ to  $v_{l} $ (Lines 18-21).  


\begin{exmp}
\label{algorithm1}
 Fig.2a shows an instance of DME-Draph. Our objective is to find a path with minimal time cost from \( v_1 \) to \( v_5 \). The weights of edges and vertices are shown in this figure.
 The time cost between two vertices is denoted as $c_i$, representing the sum of the vertex weight and the edge weight. Suppose the current time is $t_{cur} = 8:00$. 
 Figs 2b-2f illustrate the traversal of \( N(v_1) \) to compute the wait times for \( v_2 \), \( v_3 \), etc. We have  \( wt(t_{cur},w_2) = 0 \), \( wt(t_{cur},w_3) = 300 \), \( pre_2 = (v_1, 1) \), \( pre_3 = (v_1, 2) \). Therefore, \( c_2 = 180 \) and \( c_3 = 360 \).
 Next, \( v_2 \) is visited as the time cost for \( v_2 \) is the smallest. The bus line ``1'' is the same as the route with the bus line key ``1'' from $v_2$ to $v_4$, so there is no waiting time, resulting in \( c_4 = 420 \) and \( pre_4 = (v_2, 1) \). Then, we visit \( v_3 \), which has two routes leading to \( v_5 \). We compute the time costs for each route based on $wt(t_{cur},w_3)$, selecting the minimum value \( c_5 = \min \{480, 660\} \) with \( pre_5 = (v_3, 2) \). Since the time cost from \( v_4 \) to \( v_5 \) exceeds \( c_5 \), the route is determined as \( r_{min}(1,5) = \{ \langle v_1, e_{1,3,2} \rangle, \langle v_3, e_{3,5,2} \rangle \} \).
\end{exmp}

\subsection{Candidate route determination}
\label{Candidate route determination}
After determining \( r_{min}(l, l+1) \) for an OD pair \( (v_l, v_{l+1}) \), we also need to consider the dwell time $dt$ at this POI.  We update the current time to reflect the time starting from \( v_{l+1} \), specifically \( t_{cur} = t_{cur} + MC(l,l+1) + dt \). We then begin computing the next OD pair \( (v_{l+1}, v_{l+2}) \). This process continues until all OD pairs have been processed. Ultimately, the candidate route for a specific sequence \( SP_i \) is represented as $CP_i = \{ r_{min}(0,1), r_{min}(1,2), \ldots, r_{min}(n-1,n) \}$.

\begin{figure}[htbp]
\hspace{1.2cm}
    \includegraphics[width = 0.4\textwidth]{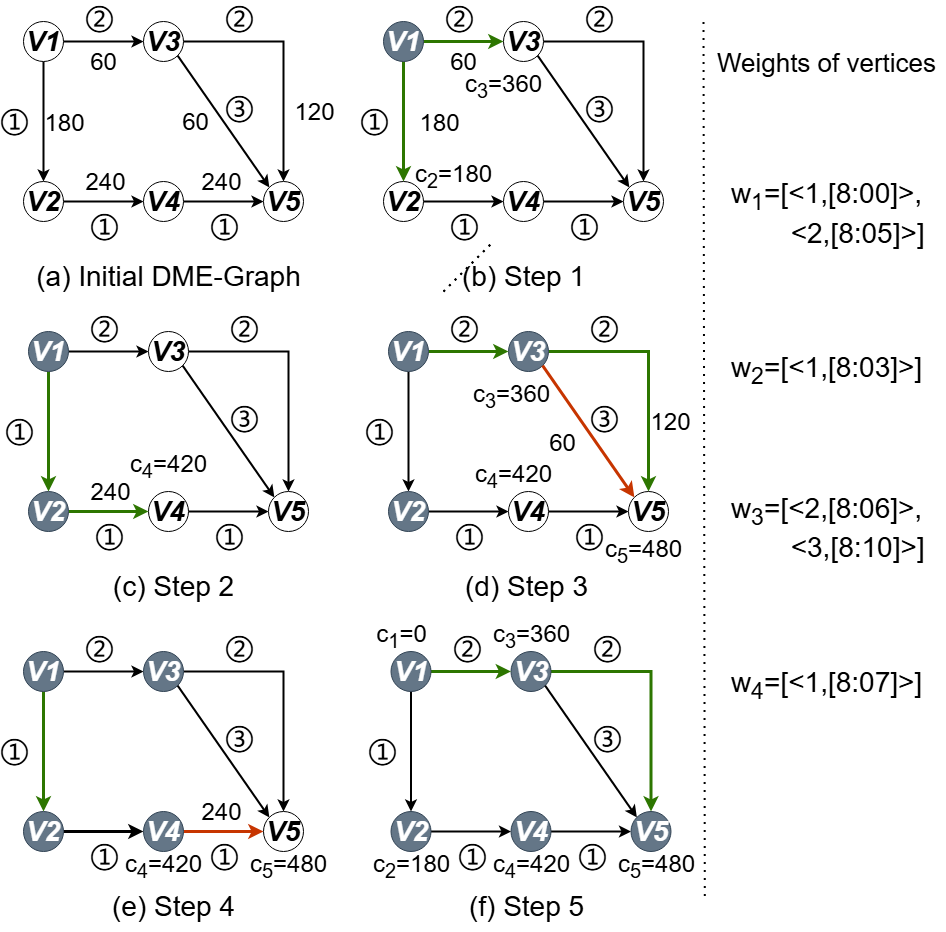}
    \caption{DSRS for Route Search}
    \label{fig:astar}
\end{figure}
 
\begin{algorithm}[htbp]  
    \footnotesize  
    \SetAlgoLined  
    \caption{DSRS Algorithm}\label{alg:DSRS}
    \KwIn{DME-Graph $DG(V, E, W)$, OD pair $(v_l,v_{l+1})$}  
    \KwOut{$r_{min}(l,l+1)$}  
    
    $c_l \gets 0$; \quad $pre_l \gets (-1,-1)$;  
    
    \For{each vertex $v_q$ in $DG$}{  
        \If{$v_q \neq v_l$}{  
            $c_q \gets \infty$; \quad Add $v_q$ into Queue;
        }  
    }  
    
    \While{Queue is not empty}{  
        $v_i \gets$ vertex in Queue with min $c_q$;  
        \If{$v_i$ is visited}{ continue }  
        Remove $v_i$ from Queue;  
        
        \For{each vertex $v_j$ in $N(v_i)$ with $e_{i,j,k}$}{  
            $x \gets w_{i,j,k} + wt(t_{cur},w_i) + c_i$;  
            \If{$x < c_j$}{  
                $c_j \gets x$; \quad $t_{cur} \gets t_{cur} + c_j$; \quad $pre_j \gets (v_i,k)$  
            }  
        }  
    }  
    
    \While{$k \neq -1$}{  
        $r_{min}(l,l+1).push(pre_{k})$; \quad $v_k \gets pre_{k}.first$; 
    }  
    
    \KwRet $r_{min}(l,l+1)$;  
\end{algorithm}

\section{EA-Star Algorithm}
\label{Sec:eastar}
The first attempt described in Section \ref{naive approach} encounters several issues: \textit{1)} For a query involving  \( n \) POIs, it computes the exact time cost for all candidate routes corresponding to \( n! \) POI visiting sequences, but most of them cannot yield the shortest route. \textit{2)} When computing the shortest route for a given a OD pair, DSRS algorithm explores the DM-Graph in all directions from the starting point, resulting in excessive and unnecessary searches.

\subsection{Basic idea}
To address these issues, we propose an enhanced heuristic search algorithm called EA-Star. This algorithm begins by estimating a lower bound for the time cost of each candidate route and ranks all candidate routes in ascending order based on these lower bounds. If the lower bound of a candidate route exceeds the minimum time cost of the already identified routes, this candidate and all subsequent ones can be discarded. For each remaining candidate route, the algorithm uses an A*-based search algorithm to improve the efficiency of identifying the shortest path between OD pairs along the route. The A*-based search algorithm predicts the minimum arrival time by calculating the ``exploration cost'' (Definition~\ref{def:explor-cost}) from the current vertex to the next POI and combines this with weights from the DM-Graph to produce an ``estimated cost'' (Definition~\ref{def:estimated-cost}). This estimated cost helps narrow the search area by highlighting the most promising directions toward the next POI.

\begin{myDef}[Exploration Cost] \label{def:explor-cost} 
For any two vertices $v_i$ and $v_j$, the exploration cost $h(i,j)$ between $v_i$ and $v_j$ is defined as the ratio of the Euclidean distance between  $v_i$ and $v_j$ to the maximum speed of the bus. 
\end{myDef}

Clearly, the actual time cost of driving from  $v_i$ to $v_j$ by taking bus must be not greater than the exploration cost between $v_i$ and $v_j$. Further, for a candidate route \(CP_i \), its exploration cost, denoted by \(h(CP_i)\), is defined as $h(CP_i) = \sum_{l=0}^{l\textless n} h(l, l + 1)$. 

\begin{myDef}[Estimated Cost]\label{def:estimated-cost}   
Each vertex $v_i$ has an estimated cost denoted by $f_i$, which consists of two parts: the actual time cost of reaching $v_i$ from $v_s$, represented by $c_i$, and the exploration cost from $v_i$ to $v_t$, represented by $h(i,t)$. Therefore, $f_i =c_i + h(i,t)$.
\end{myDef}

\subsection{Unpromising Candidate Routes Elimination}  
EA-Star introduces a pruning strategy that utilizes the exploration cost of candidate routes as a benchmark to efficiently eliminate less promising options. This strategy is based on the fact that the exploration cost is easy to calculate and always less than or equal to the actual time cost of a route. In EA-Star, we first efficiently compute the exploration cost for each candidate route. During the evaluation of the actual time cost for these routes, if no candidate routes have an exploration cost lower than the minimal time cost of the routes already identified, then the candidate route with this minimal time cost is identified as the shortest route.

The procedure is outlined as follows. First, we compute the exploration cost for each candidate route and ranking these routes in ascending order based on their exploration costs. We then process each candidate route sequentially. We use $MC_{ub}$ to denote the minimum time cost among the identified candidate routes. As each candidate route $CP_i$ is processed, its exploration cost $h(CP_i)$ is compared with $MC_{ub}$. If $h(CP_i)>MC_{ub}$, then the actual time cost of $CP_i$ must also exceed $MC_{ub}$, indicating that $CP_i$ and any subsequent routes, whose exploration costs are greater than $h(CP_i)$, cannot have a smaller time cost than $MC_{ub}$. Consequently, $CP_i$ and the subsequent candidate routes can be safely discarded. If $h(CP_i) \leq MC_{ub}$, we proceed to calculate the actual time cost of $CP_i$ and update $MC_{ub}$ if this time cost is smaller. This process is repeated until the shortest route is identified or all candidate route have been evaluated.


\subsection{Shortest route search optimization based on OD pairs}
%
For each retaining candidate route after the above pruning, EA-Star utilizes the A*-based search algorithm to enhance the identification of the shortest route between each OD pair along the route. 

For an OD pair $(v_l, v_{l+1})$ within a candidate route $CP_i$, we will use a priority queue $PQ$ to prioritize visiting the vertices with the lowest estimated costs, ensuring that each step explores the most promising direction toward $v_{l+1}$. First, we initialize the actual time cost of all vertices in the DME-Graph $DG$ to infinity and set the state of $PQ$ to empty. We then set the time cost $c_l$ to zero, and add $v_l$ to $PQ$ while computing $f_l$ (Lines 1-2). Next, we iteratively dequeue the first vertex $v_i$ from $PQ$ if $PQ$ is not empty and update the current time $t_{cur}$. If $v_i$ is $v_{l+1}$, we construct the shortest route by backtracking the arrival bus route from $v_{l+1}$ to  $v_{l}$ (Lines 3-10). Otherwise, for each adjacent unvisited vertex $v_j$ of $v_i$ ($v_j\in N(v_i)$), we compute the actual time cost of reaching $v_j$ from $v_i$ via bus line $k$, denoted as $x$ (the method for computing the actual time cost is the same as in Section~\ref{Sec:Shortest route search for OD pairs}). If the sum of $c_i$ and $x$ is less than $c_j$, it indicates that the time cost of the route from $v_l$ to $v_j$ via $v_i$ is smaller, then we update $c_j$ and $pre_j$. At the same time, we compute $f_j$ and add $v_j$ to $PQ$ (if $v_j$ is not already in $PQ$). Next, we continue to select the vertex with the lowest estimated cost from $PQ$ and repeat the process of updating the actual time costs and estimated costs for $N(v_i)$ until we reach $v_{l+1}$ or until $PQ$ is empty (Lines 11-23).

\begin{figure}[!htbp]
\hspace{0.5cm}
    \includegraphics[width = 0.43\textwidth]{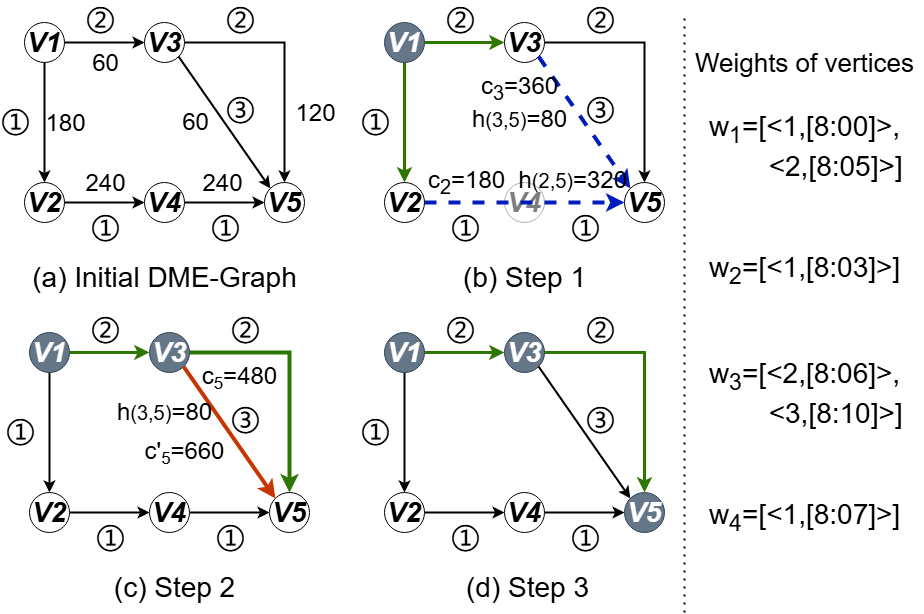}
    \caption{A*-based search algorithm for Route search}
    \label{fig:astar}
\end{figure}

\begin{exmp}
Fig. 3a shows an instance of using the A*-based search algorithm to compute the shortest route for an OD pair $(v_1,v_5)$, where the bus network is the same as that shown in Fig. 2a and the vertex weights, edge weights, and current time are the same as those in Example 2. Here, we ignore the computation of $c_i$ since its values remain the same as in Example 2. First, we traverse \( N(v_1) \) to obtain \( c_2 \) and \( c_3 \). According to Fig. 2b, the Euclidean distance from \( v_3 \) to \( v_5 \) is shorter, thus the exploration cost \( h(3,5) < h(2,5) \), leading to \( f_3 = 440 \) and \( f_2 = 500 \). We select \( v_3 \) as the vertex to visit because it has the minimum estimated cost. There are two edges from \( v_3 \) to \( v_5 \), and we choose \( e_{3,5,2} \) as it has a smaller time cost. For \( v_5 \), the estimated cost is \( f_5 = 480 \), where \( f_5 = f_3 + h(5,5) \). Therefore, we select \( v_5 \) to visit with the minimum estimated cost,thus completing the route search in Fig. 2d, expressed as \( r_{min}(1,5) = \{ \langle v_1, e_{1,3,2} \rangle, \langle v_3, e_{3,5,2} \rangle \} \).
\end{exmp}

\setcounter{AlgoLine}{0} 
\begin{algorithm}[htbp]  
    \footnotesize   
    \SetAlgoLined  
    \caption{A*-based search algorithm}  
    \KwIn{DME-Graph $DG(V, E, W)$, OD pair $(v_l,v_{l+1})$}  
    \KwOut{$r_{min}(l,l+1)$}  
    
    $PQ \gets \{v_l\}$;\quad $c_l \gets 0$;\quad $f_l \gets c_l + h(l,l+1)$\;  
    
    $pre_l \gets (-1,-1)$; 
   
    \While{$PQ$ is not empty}{  
        $v_i \gets$ vertex in $PQ$ with minimum $f_i$\;  
        
        \If{$v_i = v_{l+1}$}{  
            \While{$k \neq -1$}{  
                $r_{min}(l,l+1).push(pre_{k})$; \quad $v_k \gets pre_{k}.first$;   
            }  
            return $r_{min}(l,l+1)$\;  
        }  
        
        Remove $v_i$ from $PQ$\;  
        \quad $t_{cur} \gets t_{cur} + c_i$\;
        \For{each vertex $v_j \in N(v_i)$ with $e_{i,j,k}$}{  
            \If{$v_j$ is visited} {  
                continue  
            }  
            $x \gets w_{i,j,k} + wt(t_{cur},w_i)$\;  
            
            \If{$x + c_i < c_j$}{  
                $pre_j \gets (i,k)$\;  
                $c_j\gets x + c_i$\;
                $f_j \gets c_j+h(j,l+1)$
                \If{$v_j$ \text{ not in } $PQ$}{  
                    Add $v_j$ into $PQ$\;  
                }  
            }  
        }  
    }  
    \KwRet false\;  
\end{algorithm}

\subsection{Discussion}
\textbf{EA-Star Performance Analysis}. Given the source \( v_s \) and the destination \( v_t \) in DME-Graph, we discuss the performance of EA-Star under the following cases. \textit{(1) $h(s,t) \to 0$}. Only \( c_t \) affects priority, resulting in minimal overall impact on candidate routes as it is confined to the closest vertices;
\textit{(2) $h(s,t) \gg c_t$}. In this case, only \( h(s,t) \) influences the process, this may lead to the neglect of waiting times, ultimately resulting in the failure to search routes. To address this issue, we segment the sequence into multiple OD pairs and employ a pruning strategy to prevent excessive time costs; \textit{(3) $h(s,t) \leq c_t$}. This scenario can ensure the identification of the optimal route. However, when the difference between the two is too great, it reduces the priority disparity, ultimately causing the algorithm to degrade into a smaller range Dijkstra's algorithm, leading to an increase in possible exploration directions toward the destination. Nevertheless, it still maintains a certain level of accuracy and efficiency.

\indent\textbf{Time Complexity}. The time complexity mainly arises from the computation of candidate route searches. Let \(N\) be the number of POIs to be visited, and let \(V\) and \(E\) represent the number of vertices and edges in the DME-Graph, respectively. The naive method is similar to Dijkstra's algorithm, with a time complexity of \(O((V + E) \cdot \log V)\) for searching each OD pair's route. Therefore, the time complexity for each candidate route is \(O((V + E) \cdot \log V)\) since it consists of \(N\) OD pairs. When we explore \(N!\) candidate sequences, the final time complexity becomes \(O(N! \cdot N \cdot (V + E) \cdot \log V)\). By employing the EA-Star algorithm, the time complexity for searching OD pair routes is optimized to \(O(E)\) in the best case and \(O(V^2)\) in the worst case. Consequently, the overall time complexity is \(O(N! \cdot N \cdot E)\) in the best case and \(O(N! \cdot N \cdot V^2)\) in the worst case. Combined with the pruning strategy, this approach can significantly reduce the number of candidate routes \(N\).

\section{Experiments}
\label{Sec:expertiments}
\begin{CJK*}{UTF8}{gbsn}
\subsection{Experiment setup and datasets}
The experimental evaluation of our proposed approach was conducted using a dataset sourced from the New York City Mass Transit Authority (MTA) General Transit Feed Specification (GTFS) real-time data stream, which is publicly accessible via the official MTA website.  This dataset was meticulously filtered and restructured to create a comprehensive experimental GTFS dataset that captures travel information specifically for December weekends.
The dataset encompasses a total of 12,550 bus stops, 15,840 edges representing bus routes, and 2,525,982 recorded arrival times, thereby forming a robust foundation for establishing the bus network used in our experiments.  To assess the performance of our algorithm, we categorized the dataset into three distinct scales: small graph $SG$ with 882 vertices, medium graph $MG$ with 3,616 vertices, and large graph $LG$ with 12,550 vertices. All programs were written in C++.

In Sections \ref{6.2}-\ref{6.3}, our evaluation of the algorithm ifocuses on query time as the number of destinations grows. We will conduct multiple tests with 50 to 100 query iterations at randomly chosen stops within different scale graphs and compute the average query time. In Section \ref{6.4}, We compared the time costs of a route obtained using sequential calculations on a predefined set of POIs with those obtained using the EA-Star algorithm on a set of POIs, to demonstrate the effectiveness of EA-Star in optimizing travel time and ensuring query time stability in TSMP route planning.

\subsection{Comparison with DSRS algorithm}
\label{6.2}

We compared the DSRS algorithm with EA-Star for $SG$ and $MG$ queries, as the former is not suitable for large-scale searches, leading to excessively high query times. Fig. 4a and 4c illustrate the efficiency gains (query time ratios) of the two algorithms in $SG$ and $MG$, respectively. Fig. 4b and 4d show how the query times of EA-Star and DSRS change with the increasing number of POIs in $SG$ and $MG$. The query time of the EA-Star algorithm remains relatively stable, resulting in time optimizations ranging from several times to dozens of times. This is because, although the number of candidate sequences $n!$ grows exponentially, EA-Star performs extensive pruning on the sequences, which the naive method lacks. Additionally, EA-Star ensures that each candidate route is effectively narrowed down to the search range and ensures that the queries are more reliable.

\begin{figure}[htbp]  
    \centering  
    \begin{subfigure}[b]{0.45\linewidth}  
        \centering  
        \includegraphics[width=\linewidth]{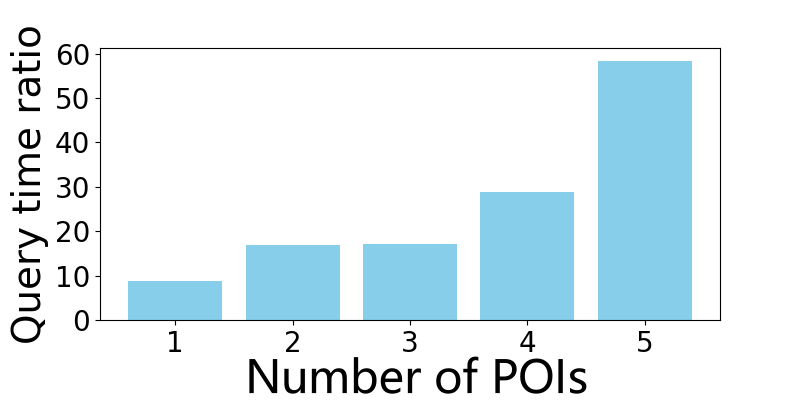}  
        \caption{Time Comparison in SG}  
    \end{subfigure}  
    \begin{subfigure}[b]{0.45\linewidth}  
        \centering  
        \includegraphics[width=\linewidth]{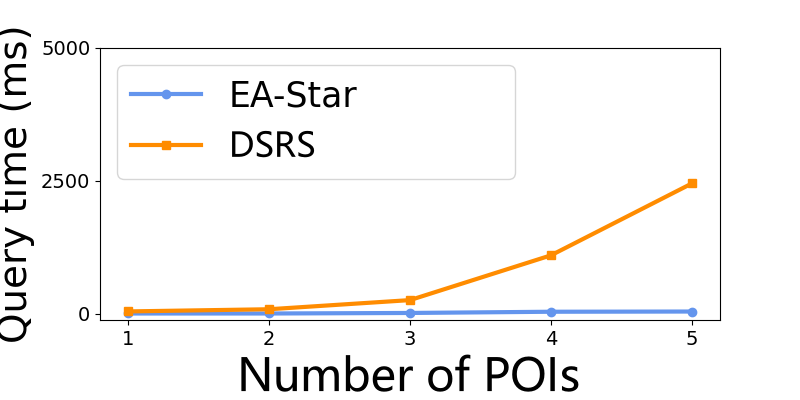}  
        \caption{Query Time in SG}  
    \end{subfigure}  
    
    \begin{subfigure}[b]{0.45\linewidth}  
        \centering  
        \includegraphics[width=\linewidth]{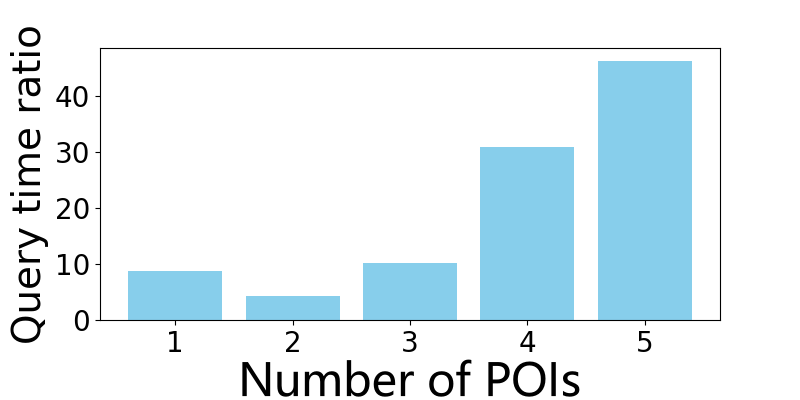}  
        \caption{Time Comparison in MG}  
    \end{subfigure}  
    \begin{subfigure}[b]{0.45\linewidth}  
        \centering  
        \includegraphics[width=\linewidth]{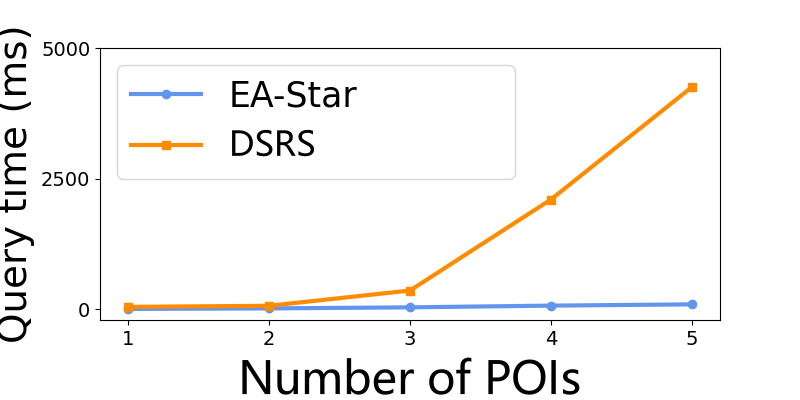}  
        \caption{Query Time in MG}  
    \end{subfigure}  
    
    \caption{Comparison of the DSRS with EA-Star}  
    \label{fig:combined}  
\end{figure}

\end{CJK*}
\vspace{-0.2cm}
\subsection{Evaluating sequence pruning}
\label{6.3}
Fig. 4a demonstrates the consistent number of pruning operations in various graphs as $n$ POIs increase, indicating the effectiveness of eliminating redundant paths. Additionally, Figs. 4b-4d present the influence of pruning on query time in different scale graphs. In cases with  POIs ($n \leq 3$), unpruning exhibits superior search efficiency due to a smaller search space and reduced overhead. However, as the number of POIs surpasses three, the advantages of pruning become more pronounced. The query time remains stable for the pruning strategy, while a sharp increase is observed in the query time for the unpruned approach. This showcases the efficiency of pruning in managing the exponential complexity growth with an increasing number of destinations and ensuring a stable and efficient search process.

\begin{figure}[htbp]  
    \centering  
    \begin{subfigure}[b]{0.45\linewidth}  
        \centering  
        \includegraphics[width=\linewidth]{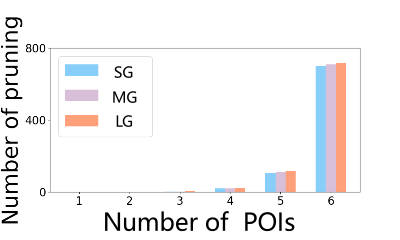}  
        \caption{SP Pruning}  
    \end{subfigure}  
    \begin{subfigure}[b]{0.45\linewidth}  
        \centering  
        \includegraphics[width=\linewidth]{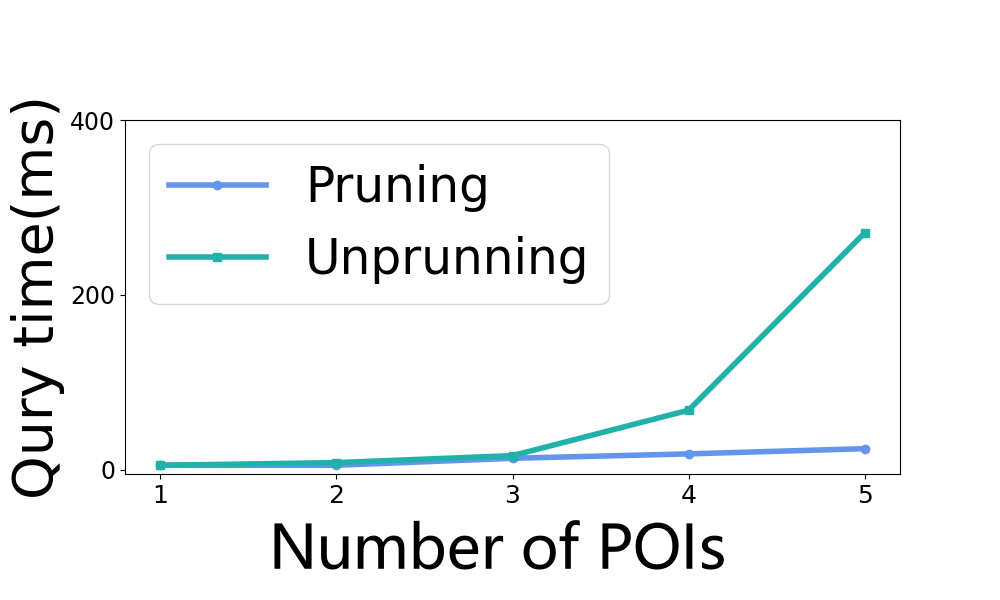}  
        \caption{Query Time in SG}  
    \end{subfigure}  
    
    \begin{subfigure}[b]{0.45\linewidth}  
        \centering  
        \includegraphics[width=\linewidth]{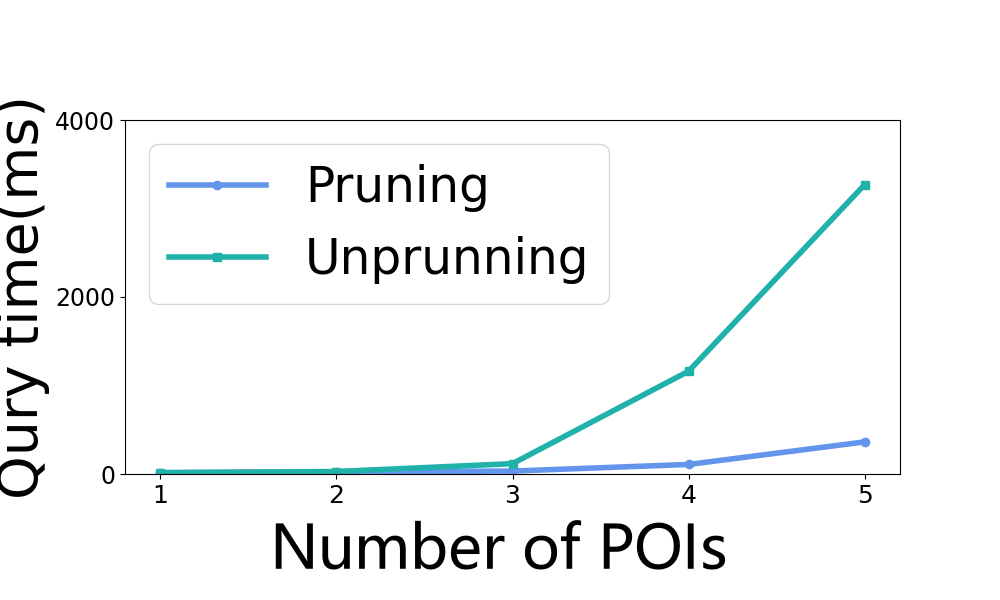}  
        \caption{Query Time in MG}  
    \end{subfigure}  
    \begin{subfigure}[b]{0.45\linewidth}  
        \centering  
        \includegraphics[width=\linewidth]{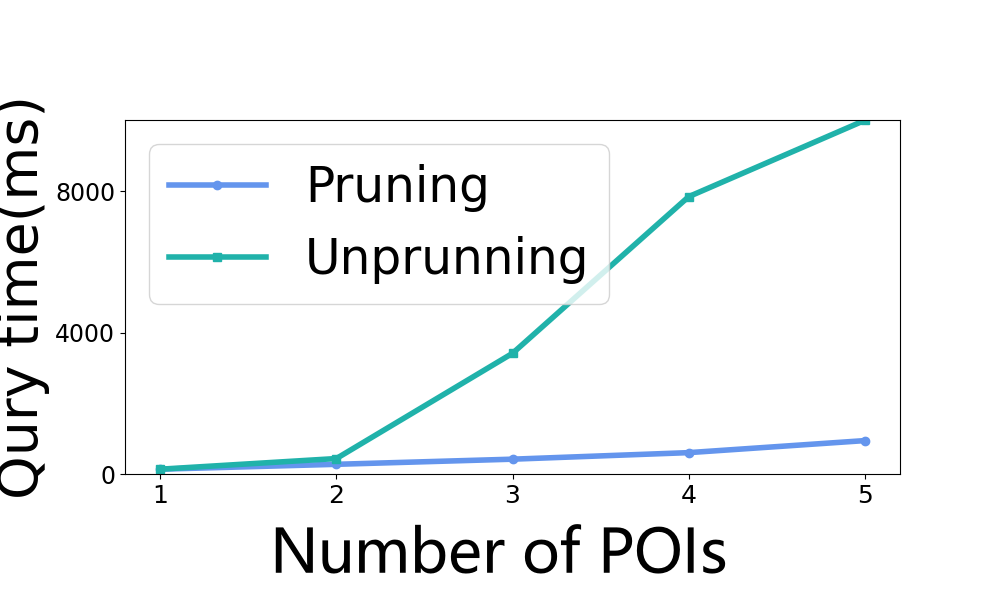}  
        \caption{Query Time in LG}  
    \end{subfigure}  
    
    \caption{Comparison for EA-Star with and without Pruning }  
    \label{fig:combined}  
\end{figure}

\subsection{Displaying EA-Star optimization effects}
\label{6.4}
We provide two TSMP route planning queries, denoted as \( Q1 \) and \( Q2 \), corresponding to two sets of POIs \( P_1 \) and \( P_2 \). After determining the optimal route for each query using the EA-Star algorithm, we compare the time costs with the routes obtained using a predetermined order of POI visits. The results demonstrate that, as the number of POIs grows, the time cost associated with the ordered sequence (i.e., visiting destinations in the order they are provided) significantly increases, while the time cost for the optimal sequence discovered by EA-Star remains relatively stable.     This trend highlights the efficiency of the EA-Star algorithm in identifying optimal paths.   The EA-Star algorithm's ability to find the optimal path leads to a substantial reduction in travel time, particularly when the query time is kept within a reasonable threshold (e.g., $n < 5$).

\begin{figure}[htbp]  
    \centering  
    \begin{subfigure}[b]{0.45\linewidth}  
        \centering  
        \includegraphics[width=\linewidth]{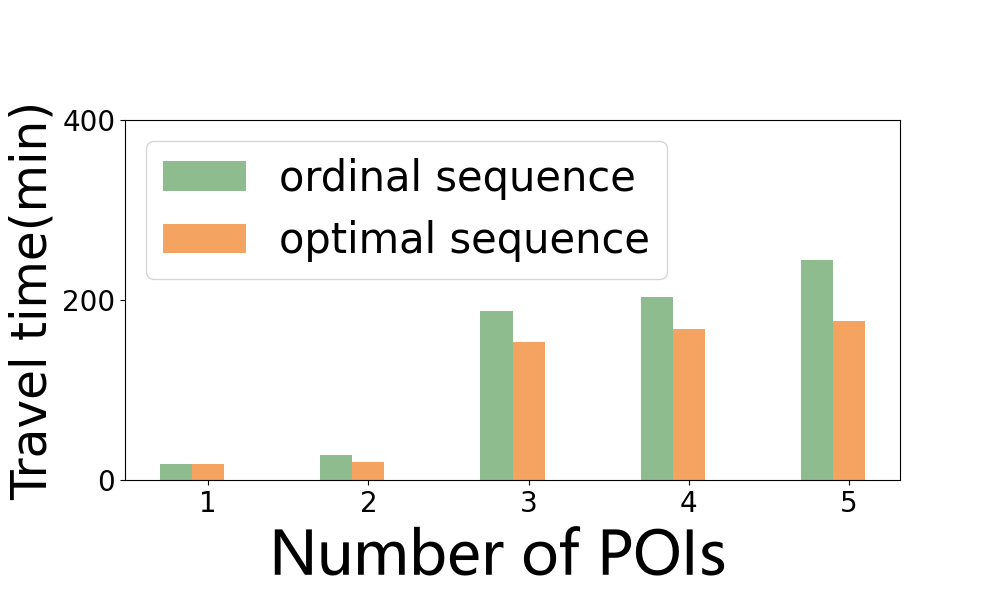}  
        \caption{Effectiveness of $Q1$}  
    \end{subfigure}  
    \begin{subfigure}[b]{0.45\linewidth}  
        \centering  
        \includegraphics[width=\linewidth]{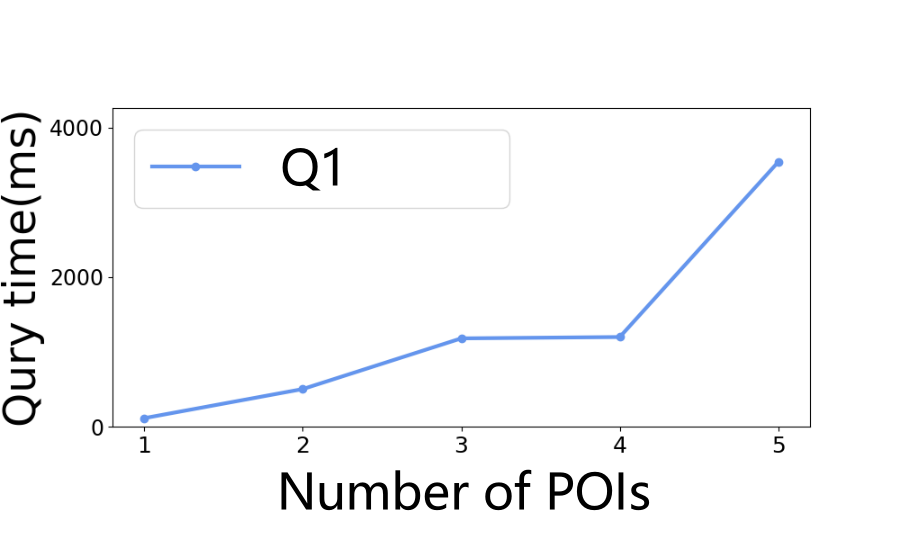}  
        \caption{Query Time of $Q1$}  
    \end{subfigure}  
    
    \begin{subfigure}[b]{0.45\linewidth}  
        \centering  
        \includegraphics[width=\linewidth]{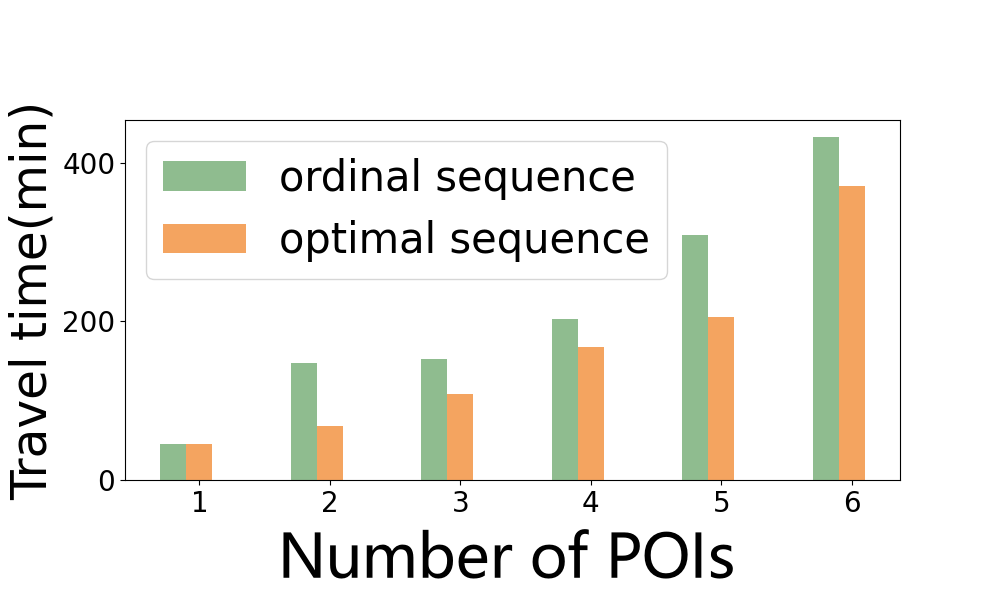}  
        \caption{Effectiveness of $Q2$}  
    \end{subfigure}  
    \begin{subfigure}[b]{0.45\linewidth}  
        \centering  
        \includegraphics[width=\linewidth]{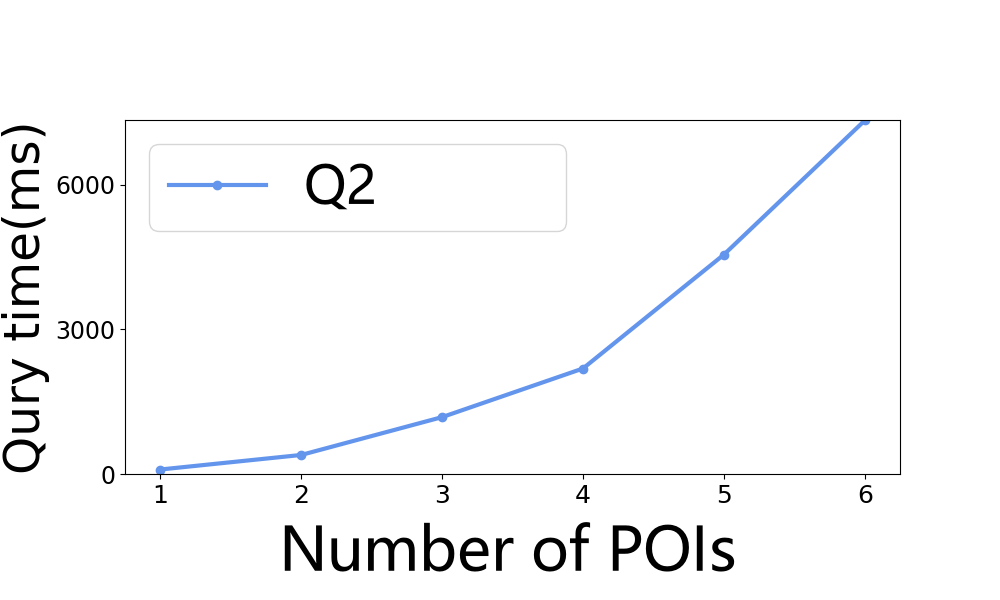}  
        \caption{Query Time of $Q2$}  
    \end{subfigure}  
 
    \caption{Query Performance of EA-Star}
    \label{fig:merged}
\end{figure}

\section{Conclusions}
\label{Sec:conclusion}
In this paper, we have addressed the intricate challenge of Time Sensitive Multi-POIs (TSMP) route planning problem on a bus network. By introducing a Directed Multiple-Edges Graph (DME-Graph), we have effectively captured the dynamic nature of bus schedules and travel times, providing a robust framework for optimizing routes.  Our proposed EA-Star algorithm, which utilizes Euclidean distance heuristics alongside an efficient sequence pruning strategy, significantly enhances the search process by narrowing down potential routes and minimizing unnecessary computations. The experimental results demonstrate the robustness and effectiveness of our approach on real-world datasets, showcasing its capability to efficiently solve unordered multi-POI bus route planning challenges while accounting for varying travel and waiting times.  The EA-Star algorithm not only improves query time stability but also ensures optimal path identification, leading to substantial reductions in travel time. In future endeavors, we aim to improve our method by integrating real-time data and dynamic updates to improve travel time prediction. Using machine learning for traffic pattern prediction and route optimization is a key method to improve accuracy. 

\section{Acknowledgements}
This work was supported by the National Science and Technology Major Project (2023ZD0121503), and by the National Natural Science Foundation of China (U23A20319, U22B2038, 62192784, 62272058, 62172351).

\bibliographystyle{plain}  
\bibliography{reference}  

\begin{thebibliography}{10}

\bibitem{r19}
Richard Bellman.
\newblock On a routing problem.
\newblock {\em Quarterly of applied mathematics}, 16(1):87--90, 1958.

\bibitem{r10}
Alican Bozyi{\u{g}}it, Gazihan Alanku{\c{s}}, and Efendi Nasibo{\u{g}}lu.
\newblock Public transport route planning: Modified dijkstra's algorithm.
\newblock In {\em 2017 International Conference on Computer Science and Engineering (UBMK)}, pages 502--505. IEEE, 2017.

\bibitem{r29}
Ade Candra, Mohammad~Andri Budiman, and Kevin Hartanto.
\newblock Dijkstra's and a-star in finding the shortest path: a tutorial.
\newblock In {\em 2020 International Conference on Data Science, Artificial Intelligence, and Business Analytics (DATABIA)}, pages 28--32. IEEE, 2020.

\bibitem{r15}
Davide Cerotti, Salvatore Distefano, Giovanni Merlino, and Antonio Puliafito.
\newblock A crowd-cooperative approach for intelligent transportation systems.
\newblock {\em IEEE Transactions on Intelligent Transportation Systems}, 18(6):1529--1539, 2016.

\bibitem{r14}
Chao Chen, Daqing Zhang, Nan Li, and Zhi-Hua Zhou.
\newblock B-planner: Planning bidirectional night bus routes using large-scale taxi gps traces.
\newblock {\em IEEE Transactions on Intelligent Transportation Systems}, 15(4):1451--1465, 2014.

\bibitem{r34}
Chao Chen, Daqing Zhang, Zhi-Hua Zhou, Nan Li, T{\"u}lin Atmaca, and Shijian Li.
\newblock B-planner: Night bus route planning using large-scale taxi gps traces.
\newblock In {\em 2013 IEEE international conference on pervasive computing and communications (PerCom)}, pages 225--233. IEEE, 2013.

\bibitem{chen2018price}
Lu~Chen, Qilu Zhong, Xiaokui Xiao, Yunjun Gao, Pengfei Jin, and Christian~S Jensen.
\newblock Price-and-time-aware dynamic ridesharing.
\newblock In {\em 2018 IEEE 34th international conference on data engineering (ICDE)}, pages 1061--1072. IEEE, 2018.

\bibitem{r25}
Yi-zhou Chen, Shi-fei Shen, Tao Chen, and Rui Yang.
\newblock Path optimization study for vehicles evacuation based on dijkstra algorithm.
\newblock {\em Procedia Engineering}, 71:159--165, 2014.

\bibitem{r18}
Edsger~W Dijkstra.
\newblock A note on two problems in connexion with graphs.
\newblock In {\em Edsger Wybe Dijkstra: His Life, Work, and Legacy}, pages 287--290. 2022.

\bibitem{ding2018ultraman}
Xin Ding, Lu~Chen, Yunjun Gao, Christian~S Jensen, and Hujun Bao.
\newblock Ultraman: A unified platform for big trajectory data management and analytics.
\newblock {\em Proceedings of the VLDB Endowment}, 11(7):787--799, 2018.

\bibitem{r31}
Franti{\v{s}}ek Ducho{\v{n}}, Andrej Babinec, Martin Kajan, Peter Be{\v{n}}o, Martin Florek, Tom{\'a}{\v{s}} Fico, and Ladislav Juri{\v{s}}ica.
\newblock Path planning with modified a star algorithm for a mobile robot.
\newblock {\em Procedia engineering}, 96:59--69, 2014.

\bibitem{r16}
Richard~E Fikes and Nils~J Nilsson.
\newblock Strips: A new approach to the application of theorem proving to problem solving.
\newblock {\em Artificial intelligence}, 2(3-4):189--208, 1971.

\bibitem{r20}
Robert~W Floyd.
\newblock Algorithm 97: shortest path.
\newblock {\em Communications of the ACM}, 5(6):345--345, 1962.

\bibitem{r30}
Andrew~V Goldberg and Chris Harrelson.
\newblock Computing the shortest path: A search meets graph theory.
\newblock In {\em SODA}, volume~5, pages 156--165, 2005.

\bibitem{r23}
John~J Grefenstette.
\newblock Optimization of control parameters for genetic algorithms.
\newblock {\em IEEE Transactions on systems, man, and cybernetics}, 16(1):122--128, 1986.

\bibitem{r8}
WANG Jian-lin.
\newblock The public transportation optimum route algorithm based on the least transfer [j].
\newblock {\em Economic Geography}, 5:673--676, 2005.

\bibitem{r38}
Chunyu Ju, Qinghua Luo, and Xiaozhen Yan.
\newblock Path planning using an improved a-star algorithm.
\newblock In {\em 2020 11th International Conference on Prognostics and System Health Management (PHM-2020 Jinan)}, pages 23--26. IEEE, 2020.

\bibitem{r28}
Natallia Kokash.
\newblock An introduction to heuristic algorithms.
\newblock {\em Department of Informatics and Telecommunications}, pages 1--8, 2005.

\bibitem{r13}
Hengyang Kuang, Yansheng Li, Yi~Zhang, and Yang Feng.
\newblock Improved a-star algorithm based on topological maps for indoor mobile robot path planning.
\newblock In {\em 2022 IEEE 6th Information Technology and Mechatronics Engineering Conference (ITOEC)}, volume~6, pages 1236--1240. IEEE, 2022.

\bibitem{r33}
Yan Li, Hongyan Zhang, Huaizhong Zhu, Jianwei Li, Wenjie Yan, and Youxi Wu.
\newblock Ibas: Index based a-star.
\newblock {\em IEEE Access}, 6:11707--11715, 2018.

\bibitem{r32}
Chenguang Liu, Qingzhou Mao, Xiumin Chu, and Shuo Xie.
\newblock An improved a-star algorithm considering water current, traffic separation and berthing for vessel path planning.
\newblock {\em Applied Sciences}, 9(6):1057, 2019.

\bibitem{r22}
Tom~V Mathew.
\newblock Genetic algorithm.
\newblock {\em Report submitted at IIT Bombay}, 53, 2012.

\bibitem{r26}
Antonio~E Mirino et~al.
\newblock Best routes selection using dijkstra and floyd-warshall algorithm.
\newblock In {\em 2017 11th International Conference on Information \& Communication Technology and System (ICTS)}, pages 155--158. IEEE, 2017.

\bibitem{r5}
Efendi Nasiboglu, Alican Bozyigit, and Yigit Diker.
\newblock Analysis and evaluation methodology for route planning applications in public transportation.
\newblock In {\em 2015 9th International Conference on Application of Information and Communication Technologies (AICT)}, pages 477--481. IEEE, 2015.

\bibitem{r24}
Anu Pradhan and G~Mahinthakumar.
\newblock Finding all-pairs shortest path for a large-scale transportation network using parallel floyd-warshall and parallel dijkstra algorithms.
\newblock {\em Journal of computing in civil engineering}, 27(3):263--273, 2013.

\bibitem{r17}
J~Russell~Stuart and Peter Norvig.
\newblock {\em Artificial intelligence: a modern approach}.
\newblock Prentice Hall, 2009.

\bibitem{r12}
Yongxin Tong, Jieying She, Bolin Ding, Libin Wang, and Lei Chen.
\newblock Online mobile micro-task allocation in spatial crowdsourcing.
\newblock 2016.

\bibitem{r35}
Zheng Wang and Jon Crowcroft.
\newblock Analysis of shortest-path routing algorithms in a dynamic network environment.
\newblock {\em ACM SIGCOMM Computer Communication Review}, 22(2):63--71, 1992.

\bibitem{r9}
Qiujin Wu and Joanna Hartley.
\newblock Using k-shortest paths algorithms to accommodate user preferences in the optimization of public transport travel.
\newblock In {\em Applications of Advanced Technologies in Transportation Engineering (2004)}, pages 181--186. 2004.

\bibitem{r6}
X~Xu, Y~Yin, and L~Liu.
\newblock Improved dijkstra’s algorithm and its application in intelligent transportation system.
\newblock {\em Journal of Residuals Science and Technology}, 13(7), 2016.

\bibitem{r36}
Yuan Yao, Zhe Peng, and Bin Xiao.
\newblock Parallel hyper-heuristic algorithm for multi-objective route planning in a smart city.
\newblock {\em IEEE Transactions on Vehicular Technology}, 67(11):10307--10318, 2018.

\bibitem{r21}
Ying~Fung Yiu, Jing Du, and Rabi Mahapatra.
\newblock Evolutionary heuristic a* search: Pathfinding algorithm with self-designed and optimized heuristic function.
\newblock {\em International Journal of Semantic Computing}, 13(01):5--23, 2019.

\bibitem{r7}
Liang Zhao and Wen-Jia Wang.
\newblock A new path search algorithm for providing paths among multiple origins and one single destination.
\newblock {\em International Journal of Computer Science and Application}, 3(1):29--33, 2014.

\bibitem{r39}
Ruicheng Zhong, Guoliang Li, Kian-Lee Tan, and Lizhu Zhou.
\newblock G-tree: An efficient index for knn search on road networks.
\newblock In {\em Proceedings of the 22nd ACM international conference on Information \& Knowledge Management}, pages 39--48, 2013.

\bibitem{r40}
Ruicheng Zhong, Guoliang Li, Kian-Lee Tan, Lizhu Zhou, and Zhiguo Gong.
\newblock G-tree: An efficient and scalable index for spatial search on road networks.
\newblock {\em IEEE Transactions on Knowledge and Data Engineering}, 27(8):2175--2189, 2015.

\end{thebibliography}

\end{document}